\documentclass[rmp]{revtex4}
\usepackage{graphicx}
\usepackage{amsmath}
\usepackage{natbib}
\usepackage{hyperref}

\begin{document}
\title{Muller's ratchet with overlapping generations}
\author{Jakob~J.~Metzger}
\affiliation{Max-Planck-Institute for Dynamics and Self-Organization, 37077 G\"ottingen, Germany}
\affiliation{Institute for Nonlinear Dynamics, Department of Physics, University of G\"ottingen, Germany}
\author{Stephan~Eule}
\affiliation{Max-Planck-Institute for Dynamics and Self-Organization, 37077 G\"ottingen, Germany}

\date{\today}
\begin{abstract}
	Muller's ratchet is a paradigmatic model for the accumulation of deleterious mutations in a population of finite size. A click of the ratchet occurs when all individuals with the least number of deleterious mutations are lost irreversibly due to a stochastic fluctuation. In spite of the simplicity of the model, a quantitative understanding of the process remains an open challenge. In contrast to previous works, we here study a Moran model of the ratchet with overlapping generations. Employing an approximation which describes the fittest individuals as
	one class and the rest as a second class, we obtain closed analytical expressions of the ratchet rate in the rare clicking regime. As a click in this regime is caused by a rare large
	fluctuation from a metastable state, we do not resort to a diffusion approximation but apply an approximation scheme which is especially well suited to describe
	extinction events from  metastable states. This method also allows for a derivation of expressions for the quasi-stationary distribution of the fittest class. 
Additionally, we confirm numerically that the formulation with overlapping generations leads to the same results as the diffusion approximation and the corresponding Wright-Fisher model with non-overlapping generations.
\end{abstract}

\maketitle

\section{Introduction}
In an asexual population of finite size, weakly deleterious mutations
can fix by chance. This phenomenon is due to stochastic fluctuations originating from the finiteness of the population, 
which can lead to a loss of the fittest class of individuals.
If the genome is assumed to be so long that back mutations are unlikely and can be ignored, the fittest
class is lost forever and the number of fixed deleterious mutations increases irreversibly. This process
has been termed Muller's ratchet \cite{Muller19642,Felsenstein:1974wq} and was observed experimentally in several studies \cite{Chao:1990bx,Duarte:1992vj,Rice:1994wt,Lynch:1996uf,Andersson:1996uo,Zeyl:2001if,Howe:2008fx}. Furthermore it has been
thought to account for the degeneration of non-recombining parts of sexually reproducing organisms such as the Y-chromosome \cite{Rice:1987wo} and mitochondrial DNA \cite{Lynch:1996uf}. Muller's ratchet can also be used to explain the absence of long-lived asexual lineages \cite{Lynch:1993tl}. 
Since in the absence of back mutations, mutation-free genomes can only be recreated by recombination between mutation-loaded classes, Muller's ratchet  provides an appealing explanation for the evolution of sex \cite{Barton:1998uq,DeVisser:2007gz}.

Each time the least-loaded class, i.e.~the class with the fewest number of deleterious mutations, is lost, it is said 
that Muller's ratchet has clicked. Since the rate of the ratchet determines the speed of degeneration of the population,
this quantity is of central interest. In its simplest form the rate of Muller's ratchet depends only on the selection coefficient $S$,
the mutation rate $U$ and the size $N$ of the population, where it is assumed that each mutation has the same effect so
individuals with $k$ mutations have fitness $(1-S)^k$. In this case the fitness space is equivalent to an axis counting the number of
deleterious mutations and the population can be organized into discrete classes labeled by the number of mutations
they carry. The deleterious mutations have the effect of shifting the population to higher values of $k$. Since the fitness of 
the respective classes is given by $(1-S)^k$, selection works into the opposite direction. In the limit of an infinitely large
population these two opposing forces lead to a steady state distribution whose precise form was found by Haigh \cite{Haigh1978251}.

If finite populations are considered, however, the calculation of the rate of Muller's ratchet turns out to be an intricate problem, 
despite its simple formulation. The difficulty arises due to the complex interaction of the fluctuation of the least-loaded class
with the rest of the distribution. A detailed quantitative understanding of the behavior of the occupation of the class with the fewest
mutations, however, is necessary to determine the mean time to extinction of this class, i.e.~the inverse of the ratchet rate. 
Despite of considerable efforts and recent advances \cite{Stephan:1993do,Gessler:1995jh,Higgs:1995dt,Gordo01032000,StephanKim2002,Etheridge:2007tv,Jain:2008uq,Waxman:2010fb,Neher01082012}, a quantitative understanding of the ratchet rate remains a challenging open problem.

In its standard form Muller's ratchet was first quantitatively described by Haigh who analyzed a classical Wright-Fisher model
of an asexually reproducing population of fixed size $N$. He pointed out that the most important quantity of the ratchet is
the average number of individuals in the least loaded class, $\bar{n}_0= N \exp(-U/S)$, because fluctuations of $n_0$ ultimately
lead to a click of the ratchet. Later it was shown by Jain that the ratchet rate cannot depend only on $\bar{n}_0$ but rather has to depend
on $\bar{n}_0 S$ \cite{Jain:2008uq}. If $\bar{n}_0 S$ is small, the ratchet clicks frequently and the populations behaves like a wave in $k-$space propagating towards
higher values of $k$. The traveling wave approach to Muller's ratchet  was discussed in \cite{Rouzine200824} and provides an appealing
quantitative theory for frequently clicking ratchets. 

While this regime of Muller's ratchet is relatively well understood, a quantitative understanding of the opposite case $\bar{n}_0 S\gg 1$ of a rarely clicking ratchet is still lacking
and has recently attracted a lot of attention \cite{Jain:2008uq,Neher01082012}. In this regime the rate of the ratchet is exponentially small in $\bar{n}_0 S$ \cite{Jain:2008uq} and extinction of the fittest class occurs as the result of a rare large fluctuation. In contrast to the fast clicking regime the distribution of the population equilibrates to a metastable state after each click. A wide-spread approach in this regime is therefore to consider only the fittest class and apply a phenomenological model for all the classes $n_k$, $k>0$, with more mutations than the least-loaded class. Such an approach leads to a one-dimensional approximation where just the fittest class is taken into account. Generally the rate of the ratchet can then be calculated by means of a diffusion approximation
as the result of a one-dimensional mean-first passage problem. Recently it was shown how this approach can be improved by
accounting for the interaction of the fluctuations of the fittest class with the tail of smaller fitness which can lead to a
delayed feedback \cite{Neher01082012}.

Up to now a quantitative treatment of Muller's ratchet relied either on Haigh's model or on the corresponding diffusion
approximation.  To our knowledge a Moran formulation with overlapping generations has not been employed so far.
This is not surprising as in the diffusive limit any quantity should become independent of the respective microscopic formulation
and a Moran formulation of Muller's ratchet is expected to be computationally disadvantageous.
A Moran formulation, however,  can also lead to interesting new approaches to tackle the problem of the ratchet rate analytically.

In the present work we investigate a Moran formulation of Muller's ratchet and show how this model can
be approximated by a one-dimensional Moran-process in the regime $\bar{n}_0 S\gg 1$ where the
ratchet clicks infrequently. We show that this model allows for an analytical solution for the ratchet rate
which agrees almost perfectly with values obtained by numerical simulations of the full ratchet. 
Furthermore, by employing a recently developed method to treat rare
large fluctuations in stochastic population dynamics, we find analytical expressions for the ratchet rate and the quasi-stationary distribution of the fittest class in the parameter range $U/S\leq 2$ which also agree very well with the corresponding results of the full ratchet. Finally, we confirm numerically that the formulation with overlapping generations leads to the same results as the diffusion approximation and the corresponding Wright-Fisher model with non-overlapping generations.

\section{Models and Methods}
\label{sec:Models_and_Methods}
In the standard formulation of Muller's ratchet, as considered by Haigh \cite{Haigh1978251}, mutations in a population of fixed size $N$ occur at rate $U$ and individuals are classified into different groups according to the number of deleterious mutations they carry, $k$. Each mutation reduces the fitness of the genotype $k$ by an amount $S$ such that the growth rate of an individual with $k$ mutations is proportional to $(1-S)^k$. The  ratio of the mutation rate $U$ to the mutation effect $S$, 
which is denoted by $\Lambda = U/S$, plays a central in the analysis of the ratchet.

The reproduction model usually employed in the analysis of Muller's ratchet is Wright-Fisher sampling. It consists of, at each time step, replacing the whole generation of individuals by a multinomial resampling of the current generation  \cite{blythe_stochastic_2007} weighted by the fitness of the different classes. Thus, according to Haigh \cite{Haigh1978251}, if $n_k(t)$ is the number of individuals in generation t which carry $k$ mutations and $\mathbf{n}(t) = (n_0, n_1, ...)$, then the distribution of $\mathbf{n}(t+1)$ is multinomial with parameters $N$ and $\{p_k(t), k = 0,1,...\}$, where 
\begin{equation}\label{Haighmodel}
p_k(t) = \frac{1}{\bar{W}} \sum_{j=0}^k n_{k-j}(t)(1-S)^{k-j} e^{-\Lambda}  \frac{\Lambda^j}{j!}
\end{equation}
and the mean fitness $\bar{W}$ is given by $\bar{W} = \sum_{j=0}^\infty (1-S)^j p_i$.

Wright-Fisher sampling has the advantage of being very efficient for numerical simulations. The downside of the model
is, however, that it does not easily allow for analytical methods to be used. Therefore, the corresponding diffusion approximation of the microscopic Wright-Fisher formulation is usually used to predict the click rates of the ratchet.

The second widely applied reproduction model in population genetics is the Moran process which in contrast to the
Wright-Fisher formulation assumes overlapping generations. 
The Moran process on which we focus in this article is amenable to a wider range of analytical methods (at the cost of being slower in numerical simulations) \cite{park_speed_2010}. It is a stochastic process in which at each time step one individual is chosen for reproduction and one for removal from the population. The choice of the individual that reproduces is random, but (similarly to the Wright-Fisher formulation) weighted by the fitness of the class the individual is chosen from. The probability of removal (or death) of an individual is independent of the fitness. Applied to Muller's ratchet this therefore embodies the following procedure: An individual with $k$ mutations is chosen according to the abundance and selection preference of the class $k$ with weight $(1-S)^k n_k/ \sum_{j=0}^\infty (1-S)^j n_j$. This individual spawns one offspring with $k$ mutations that can then mutate to $k+l$ mutations with probability $e^{-U}\, U^l /l!$. The probability to mutate is thus $1-e^{-U}$, which is the same as in the Wright-Fisher model. Also, one individual with $k$ mutations is chosen for removal with probability $n_k/N$ (this may be the one that reproduced). Since on average every individual is chosen for removal once every N time steps, it is natural to define one generation in the Moran model as N time steps. In all figures, the ratchet click times are thus expressed in generations.

Although different on the microscopic scale, both Wright-Fisher and Moran models usually converge to the same mesoscopic diffusion regime when $N$ is large and fitness advantages and mutation rates are of order $N^{-1}$. In this limit, the equation describing the evolution of the population is given by
\begin{equation}
	\frac{d}{dt}n_k = S(\bar{k}-k)n_k - U\,n_k +U\,n_{k-1} +\sqrt{n_k}\eta_k \label{eq:diff}
\end{equation}
where $\bar{k} = N^{-1}\sum_k k\,n_k$ \cite{Neher01082012}. The uncorrelated Gaussian white noise $\eta$ with $\langle \eta_k(t) \eta_l(t') \rangle = \delta_{kl}\delta(t-t')$ models the stochastic fluctuations due to the finiteness of the population (genetic drift). In the infinite population limit, this equation becomes deterministic and has the steady-state solution $\bar{n}_k=N e^{-\Lambda} \Lambda^{k}/k!$ \cite{Haigh1978251}. Also, a time dependent solution of the deterministic model has been obtained \cite{Etheridge:2007tv}. In this paper, we solve Eq.~\eqref{eq:diff} numerically using stochastic Runge-Kutta methods \cite{Roessler:2010:RMS:1958393.1958409}.

\section{Approximative one-dimensional Moran model of Muller's ratchet}
\begin{figure}[htbp]
	\centering
		\includegraphics[height=3in]{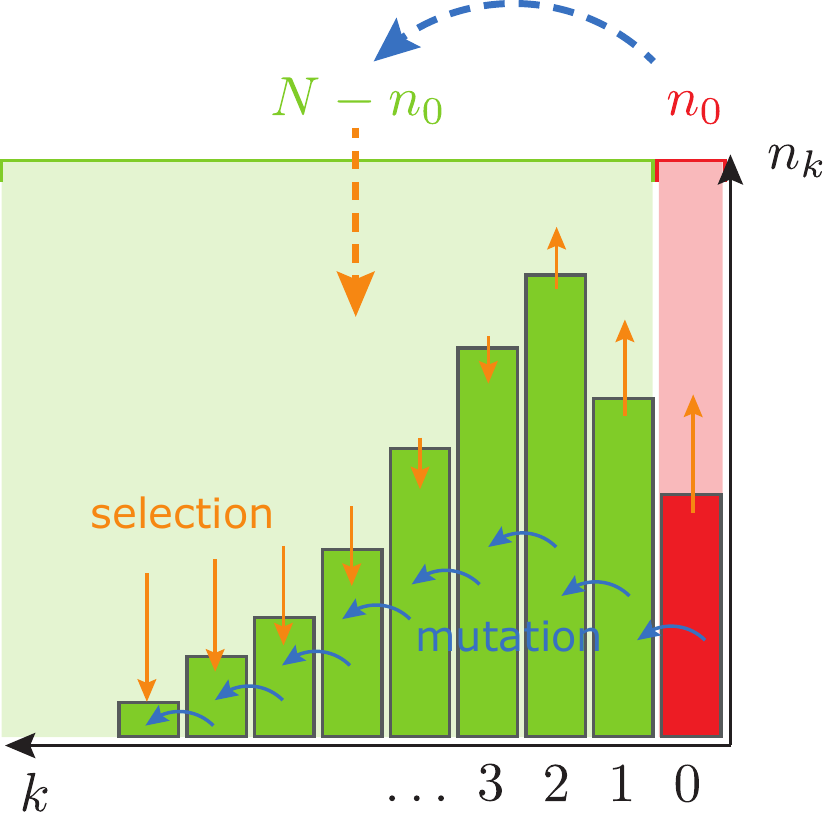}
	\caption{Illustration of Muller's ratchet in the space of deleterious mutations. Individuals are grouped into different classes depending on the number of mutations $k$ they carry. Mutation (blue arrows) drives the population to higher values of $k$, while selection (yellow arrows) opposes this motion, leading to a quasi-stationary distribution. The two-state approximation amounts to putting all mutated individuals in one mutated class (light green box). Both mutation into this class and selection pressure operating on it (large arrows) have to be calculated from the original mutation rates and selection strengths. Since the total population size is conserved, calculating the distribution of the number of individuals in the two classes reduces to the analysis of $n_0$ and thus to a one-dimensional problem.}
	\label{fig:fig_ratchet}
\end{figure}

A mathematical analysis of the Moran model for Muller's ratchet is complicated and
even the formulation of the corresponding Markov chain \cite{ewens2004mathematical} is involved and barely leads to new insights. 
The important advantage of Moran models, however, is that they can be formulated in terms of a master equation \cite{gardiner2009stochastic}. 
Therefore Moran models are analytically tractable even beyond the diffusion approximation, if only two species are considered. 
Thus an appealing approach to the analysis of the rate of Muller's ratchet is to approximate the full ratchet by a model
consisting only of two species. Since we are interested in the loss of the fittest class with zero mutations a natural choice is to consider individuals with zero mutations as one species, and to combine all others in a 
class which contains all individuals with one or more mutations. The constraint of a fixed population size $N$ then
leads to a one-dimensional model. An illustration of the distribution of individuals in the space of mutations and the reduction to a one-dimensional model is given in Fig.~\ref{fig:fig_ratchet}

The master equation for the probability $P(n, t)$ to find $n=n_0$ individuals in the fittest class is
\begin{equation}\label{master}
\frac{dP(n,t)}{dt}=T_-(n+1)P(n+1, t)+T_+(n-1)P(n-1, t)-(T_+(n)+T_-(n))P(n, t) \, ,
\end{equation}
with the transition rates
\begin{align}	
	T_+(n) = T(n+1|n) &= (1-u)\frac{(1-\frac{n}{N})\frac{n}{N}}{1-s(1-\frac{n}{N})}\label{eq:mor_2rates1}\\ 
	T_- (n)= T(n-1|n) &= (1-u)\frac{(1-s)(1-\frac{n}{N})\frac{n}{N}}{1-s(1-\frac{n}{N})} + u\frac{n}{N}\label{eq:mor_2rates2}
\end{align}
where $u$ is the mutation rate away from the fittest class and $s$ is the fitness disadvantage when carrying
a mutation \cite{blythe_stochastic_2007}. (Here and in the following lowercase letters denote the parameter values of the approximative
two-state model, while capital letters denote the parameters of the full ratchet.)
The idea of representing all classes but the fittest as one class was first introduced in \cite{Waxman:2010fb} 
for a Wright-Fisher model of Muller's ratchet.

A crucial step in the reduction of the full model of Muller's ratchet to the one-dimensional formulation 
is the relation of the two mutation rates and fitness disadvantages in the respective models. 
This mapping is not unique and two reasonable assumptions have to be invoked to relate the two parameters
pairs. A plausible approach is to compare the steady state distributions in the infinite population limit of the respective models.
For the full ratchet the well-known steady state distribution for the probability of an individual to have $k$ mutations 
is $p_s(k)=e^{-U/S}(U/S)^k / k!$. A non-zero steady state of the fittest class 
in the two-state system can only be obtained in the parameter regime $u/s<1$ and is given by $n^*= N(1-u/s)$. To relate the parameters we now demand that $(i)$ the mean fitness of the full population and $(ii)$ the mean fitness of all individuals carrying a mutation is equal in both models. The mean fitness of the full population in the steady state of the full ratchet is  $\sum_{k=0}^\infty p(k) (1-S)^k = e^{-U}$ while the mean fitness of all individuals in the two state model is $N^{-1}(n^*+(1-s)(1-n^*))=1-u$. 
Condition ($i$) accordingly suggests the relation
\begin{equation}\label{relation1}
u=1-e^{-U}\, .
\end{equation}
The mean fitness of all individuals carrying mutations is in the full ratchet model given by $\sum_{k=1}^\infty p(k) (1-S)^k = e^{-U} - e^{-U/S}$.
In the two state model this corresponds to $(1-s)\,(N-n^*)/N=(1-s)u/s$. Employing condition $(ii)$ consequently yields the relation 
\begin{equation}\label{relation2}
s=\frac{1-e^{-U}}{1-e^{-U/S}}\, .
\end{equation}
We can also introduce the parameter $\lambda=u/s$ which is related to $\Lambda=U/S$ according to
\begin{equation}\label{relation3}
\lambda=1-e^{-\Lambda}
\end{equation}
Relation (\ref{relation3}) shows that the restriction $\lambda<1$ of the two-state model does not restrict
the range of $\Lambda$.

Before we present the analytical solution for the ratchet rate of this model, let us shortly discuss the validity of the 
approximation used. To correlate the parameters of the full ratchet and the two state model, we have related properties of the 
equilibrium solution of an infinite population in both models. This certainly makes sense as long as the typical time $t_r$
that it takes for the population to relax to a metastable state after each click is much smaller than the mean time $\tau$ 
between two successive clicks. This condition is fulfilled in the case of the slowly clicking ratchet, which is the regime we focus on in this work. If the ratchet clicks rapidly the population does not equilibrate after a click and relating the parameters based on equilibrium distributions is clearly not valid.

A consequence of our relation (\ref{relation1}) is that the mutation rates out of the fittest class are equal in both models
which certainly is a reasonable assumption \cite{Waxman:2010fb}. Furthermore our second relation (\ref{relation2}) entails that the number of individuals which are
not in the fittest class is the same in the equilibrium states of both models, i.e.~$\bar{n}_0=n^*$. Consequently the same holds true for the number of individuals in the fittest class. Thus, although the parameter mapping is not unique, it is hard to think of any other relation in the slowly clicking regime as this would consequently violate the properties specified above. It is important to keep in mind
that the parameter mapping is only valid in the rare clicking regime and that other mappings might be more appropriate in the fast clicking regime \cite{Waxman:2010fb}.

\subsection{Exact solution and Comparison with Full Moran ratchet}
With the reduction to a two-state problem as given in the previous section, we can now exploit the advantages that the Moran formulation offers for analytical calculations. The mean click time of the ratchet is given by the mean first time of the population with no mutations reaching zero. It is well known that a solution of such a mean first passage time problem in a two-state model can be formally written as a product of the transitions rates \eqref{eq:mor_2rates1} and \eqref{eq:mor_2rates2} and is given by \cite{gardiner2009stochastic}
\begin{equation}
	\tau = \sum_{y=1}^{n_i} \phi(y) \sum_{z=y}^{N-1} 1/(T_+(z) \phi(z)) \label{eq:tauMA}
\end{equation}
where $\phi(n) = \prod_{k=1}^n T_-(k)/T_+(k) $ and $n_i$ is the initial number of individuals in the lowest mutation class population. This expression can be evaluated for moderate $N$, but the number of terms grows quickly with $N$ which makes it more and more difficult to evaluate $\tau$.

\begin{figure}[htbp]
	\centering
		\includegraphics[height=3in]{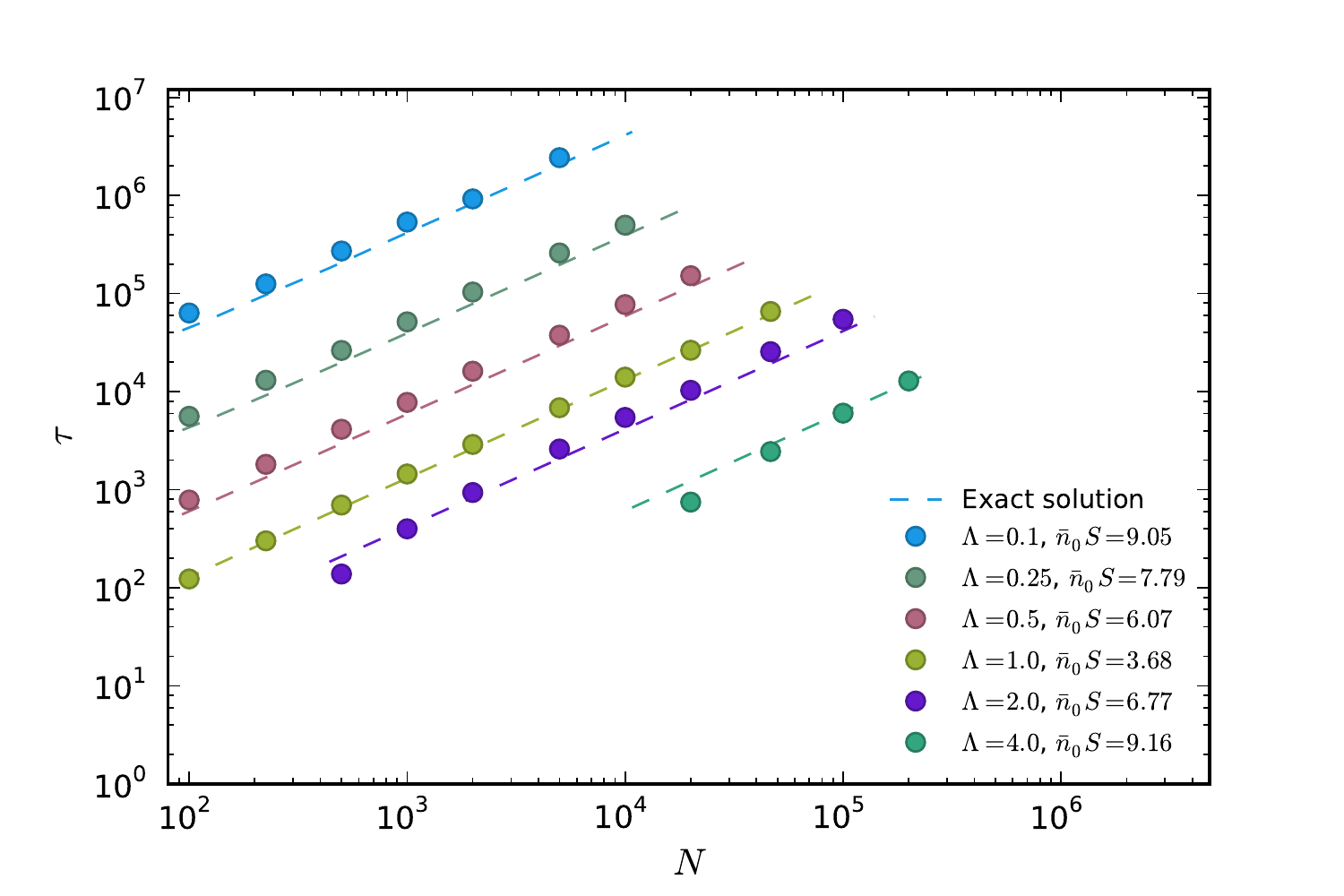}
	\caption{Comparison of the analytical expression of the mean click time (i.e.~the inverse ratchet rate) for the two-state model and numerical simulations of the full ratchet in the slow ratchet regime. Different colors correspond to different effective mutation rates $\Lambda = U/S$ and different $\bar{n}_0 S = N S e^{-\Lambda}$. }
	\label{fig:Moran_compare}
\end{figure}

To compare our analytical results to the full ratchet we have performed extensive numerical simulations of the full Moran ratchet using the rules detailed in section \ref{sec:Models_and_Methods}. We organize our results as follows: The parameters $U$ and $S$ are grouped according to the conditions specified below, and then $N$ is varied. Since the selection penalty $S$ can be interpreted as a timescale \cite{Neher01082012}, we group parameters with the same rescaled mutation rate $\Lambda = U/S$. Similarly, since $(N\,S)^{-1}$ can be interpreted as rescaled variance of the stochastic effects \cite{Neher01082012}, we also group parameters with the same $N\,S$, which is then equivalent to keeping $\bar{n}_0\,S = N S e^{-\Lambda}$ fixed. The corresponding two-state parameters are rescaled as given by \eqref{relation1} and \eqref{relation2}.
A comparison of the analytical results and the simulations is given in Fig.~\ref{fig:Moran_compare}. We observe excellent agreement of the analytic result given by Eq.~\eqref{eq:tauMA} for the two-state model with the simulation of the full ratchet
in the slow ratchet regime, where the two-state approximation is valid.

\section{WKB-approximation to the ratchet rate}
The expression (\ref{eq:tauMA}) for the mean time to extinction is exact. It gets, however, unwieldy and impractical when
larger population sizes are considered. Furthermore it does not allow for any analytical statements about the
distribution of the frequency of the fittest class. Therefore, we want to gain quantitative insight into the ratchet rate
and the distribution of the frequency of the fittest class by an approximative treatment of Eq. \eqref{master}. 
The most widely applied approach certainly is the diffusion
approximation from which by standard methods the mean time to extinction (MTE) $\tau$ 
can be calculated analytically \cite{gardiner2009stochastic}. The resulting expression usually has to be evaluated numerically. 
While the diffusion approximation provides faithful results in the regime
where an extinction event is the outcome of a typical fluctuation of the process, it in general may fail to describe
the MTE correctly when extinction occurs as the result of a rare large fluctuation \cite{Ovaskainen:2010cl,Doering:2013bd,Hanggi:1984bo}.
In the rare clicking regime the relaxation time to the metastable state is much shorter than the mean time between the clicks 
and the population equilibrates after each click. It is important to note that in such a scenario the click of the ratchet is due to a rare large fluctuation away from the metastable state. 

An approach to the treatment of master equations which is especially well suited to account for rare event statistics
is the WKB- (Wentzel-Kramers-Brillouin) theory. This approximation scheme which is sometimes
referred to as the eikonal approximation was first developed for a semi-classical treatment
of quantum mechanics and has recently attracted a lot of attention in the context of stochastic population dynamics
 \cite{Dykman:2008ig,Schwartz:2011uy,Assaf:2011iw,Black:2011ep}. Similar to the diffusion approximation, it replaces the master equation of the Moran process by an analytically tractable
equation which in addition allows for a mathematically controlled approximation in terms of powers of the inverse
population size. Recently the WKB-approximation has also found its way into evolutionary modeling \cite{Ovaskainen:2010cl,Assaf:2010de,Assaf:2011hs,Black:2012dz}.
The approach we apply in the following was first considered in \cite{Dykman:1994hw} and later considerably 
extended and generalized in \cite{Assaf:2010de}.

The basic idea relies on the fact that the process can be characterized by
a metastable state around which the frequency of the fittest class resides. After a long average time
$\tau$ the fittest class is eventually lost and Muller's ratchet clicks. For the approach to work two
crucial assumptions have to be made. First, the population size has to be finite and not too
small, i.e.~$N\gg 1$. Second the typical relaxation time $t_r$ to the metastable state
should be much shorter than the MTE, i.e.~$t_r\ll \tau$. We note that here this condition has to
hold anyway in order for the two state approximation to be meaningful.
It can be shown that the metastable state, which is sharply peaked around 
$n^*$, is encoded in the first excited eigenvector $\Pi(n)$ of the master equation (\ref{master}) 
which has not decayed at a time $t\geq t_r$ \cite{Assaf:2010de}. 
Thus the shape of the PDF of the metastable state, which is referred to as the quasi-stationary distribution (QSD),
is given by $\Pi(n)$. Furthermore, the decay rate of this distribution, i.e.~the ratchet rate $\tau^{-1}$, is determined
by the first non-zero eigenvalue of the master equation (\ref{master}). As was shown in \cite{Assaf:2006ii}, the decay of the 
QSD for times $t\gg t_r$ can therefore be obtained as
\begin{equation}\label{QSDdecay}
P(n>0, t)=\Pi(n) e^{-t/\tau}\, .
\end{equation}
 Accordingly, the click probability distribution behaves as
 \begin{equation}\label{clickpdfgrowth}
 P(0, t)= 1-e^{-t/\tau}\, .
 \end{equation}
Using Eqs.(\ref{master}), (\ref{QSDdecay}) and (\ref{clickpdfgrowth}) the click rate is given by
\begin{equation}
\tau^{-1}=T_-\left(n=1\right)\Pi\left(n=1\right)\, ,
\end{equation}
which is just to the probability flux into the absorbing state $n=0$. 

In following we present an approximative approach to calculate the QSD $\Pi(n)$ based on a WKB-type approximation.
Inserting (\ref{QSDdecay}) into (\ref{master}) one obtains after introducing $x=n/N$ 
\begin{equation}
\frac{\pi(x)}{\tau}=t_-\left(x+\frac{1}{N}\right)\pi\left(x+\frac{1}{N}\right)+t_+\left(x-\frac{1}{N}\right)\pi\left(x-\frac{1}{N}\right) -\left[t_+(x)+t_-(x)\right]\pi(x)\, ,
\end{equation}
where $\Pi(n)=\Pi(Nx)=\pi(x)$ and $T_\pm(n)=T_\pm(Nx)=t_\pm(x)$.
Since we consider the rare-clicking regime of the ratchet, the term on the left-hand side is exponentially
small in $N$ and we can neglect it. The resulting quasi-stationary master equations reads 
\begin{equation}\label{QSDmaster}
0=t_-\left(x+\frac{1}{N}\right)\pi\left(x+\frac{1}{N}\right)+t_+\left(x-\frac{1}{N}\right)\pi\left(x-\frac{1}{N}\right) -\left[t_+(x)+t_-(x)\right]\pi(x)\, .
\end{equation}
Now we are ready to employ the WKB approach by expressing the solution of this equation by the ansatz \cite{Dykman:1994hw}
\begin{eqnarray}
\pi(x) & = & A(x) \exp \left(-N \left[S_0(x) + \mathcal{O}(N^{-1})\right]\right) \nonumber \\
         & = & C \exp (-N S_0(x)- S_1(x)+ \mathcal{O}(N^{-1})) \label{WKBansatz}
\end{eqnarray}
where both $S_0(x)$ and $S_1(x)$ are assumed to be of order unity and $C$ is a normalization constant.
Inserting this ansatz into (\ref{QSDmaster}), expanding $S_0(x+1/N)$ around $x$ to first order
and neglecting terms of order $\mathcal{O}(1/N)$, we obtain in leading order
\begin{equation}\label{Hamilton}
0=t_+(x)\left(e^{S'_0(x)}-1\right)+t_-(x)\left(e^{S'_0(x)}-1\right)\, ,
\end{equation}
where $S'_0(x)=\frac{d}{dx} S_0(x)$. The solution of this equation is given by
\begin{equation}
S_0(x)=\int^x \ln\left[ \frac{t_+(y)}{t_-(y)} \right] dy\, .
\end{equation}
After insertion of $S_0(x)$ into the ansatz (\ref{WKBansatz})  the lowest order solution for the
QSD is obtained up to the normalization constant $C$. To determine $C$ one exploits the fact that the QSD is strongly peaked
around $x^*=n^*/N$ and then assumes it to be of Gaussian shape centered at $x^*$ which is normalized
to unity. Around the maximum $x\simeq x^*$ this leads to an approximation of the QSD by 
$\pi(x)\simeq C e^{-NS_0(x^*)-(N/2) S_0''(x^*) (x-x^*)^2}$ whose normalization yields $C= e^{-NS_0(x^*)}$.
Hence in leading order we obtain for the QSD
\begin{equation}
\pi(x)\sim e^{N[S_0(x^*)-S_0(x)]}\, .
\end{equation}
Using this expression of the QSD we can calculate the leading order behavior of the click rate
\begin{equation}
\tau^{-1}\sim e^{-N[S_0(x^*)-S_0(0)]}\, ,
\end{equation}
where we have used that $\pi(N^{-1})\sim\pi(0)$ and $t_-(N^{-1})\sim t'_-(0)/N$ for large $N$. In leading order we thus obtain
the anticipated exponential dependence of the ratchet on $N$ in the rare clicking regime. These results are 
valid as long as $N[S_0(x^*)-S_0(0)\gg 1$ because the WKB-ansatz requires the ratchet rate to be exponentially small.
Furthermore the normalization procedure can be expected to fail if the metastable state is close to boundary $x^*\simeq 1$
because the Gaussian approximation does not hold anymore.

The next order $\mathcal{O}(1)$-corrections of the WKB-approximation 
are obtained by expanding $S_0(x+1/N)$ to second order and $S_1(x+1/N)$ to first order around $x$
and provide the pre-factor of the QSD and the ratchet rate. The calculation 
of the sub-leading corrections is more involved and shall not be carried out in detail here. 
The crucial step in the calculation is to note that the WKB-solution in leading order is not valid close 
to the absorbing state at $x=0$. 
Therefore the WKB solution has to be matched with an exact recursion solution of quasi-stationary master equation (\ref{QSDmaster}). A detailed account of this method is given in \cite{Assaf:2010de}.
Following the steps in \cite{Assaf:2010de} we obtain for the QSD
\begin{equation}\label{QSDfullsol}
\pi(x)=\frac{\sqrt{S''(x^*)}t_+(x^*)}{\sqrt{2\pi N t_+(x)t_-(x)}}e^{N[S_0(x^*)-S_0(x)]}\, .
\end{equation}
The WKB solution for the inverse of the ratchet rate is given by
\begin{equation}
\tau=\frac{\sqrt{2\pi N R}}{\sqrt{S''(x^*)}t_+(x^*)(R-1)}e^{N[S_0(x^*)-S_0(0)]}
\end{equation}
with $R=t'_-(0)/t'_+(0)=(u-1)/(s-1)$.

\begin{figure}[htbp]
	\centering
		\includegraphics[height=3in]{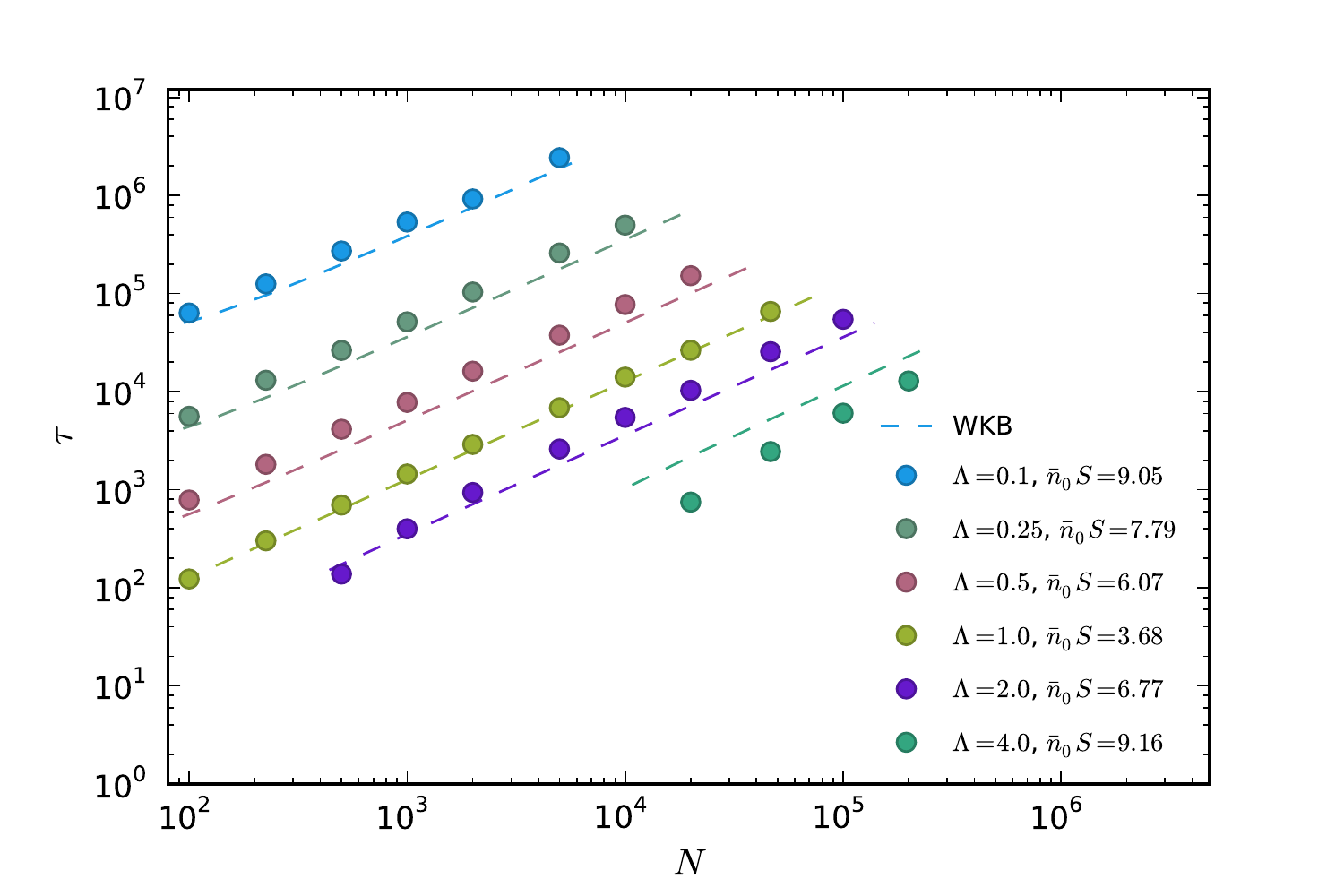}
	\caption{Comparison of the WKB-solution (\ref{tausolfull}) for the inverse ratchet rate and the numerical simulations of the full ratchet. Different colors correspond to different effective mutation rates $\Lambda = U/S$ and different $\bar{n}_0 S= N S e^{-\Lambda}$}
	\label{fig:WKB_compare}
\end{figure}

Inserting the respective transition rates, we obtain for the mean time to extinction of the fittest class, i.e.~the inverse of the ratchet rate

\begin{equation}\label{tausolfull}
 \tau= \frac{N s (1-u) (N (s-1)-s) \left(u/s (1-s)^{-\frac{1-s}{s+u-1}} \left(u/s(1-u)
	   \right)^{\frac{1-s}{s+u-1}}\right)^{-N}} {\sqrt{u/(2\pi N(1-s))} (s-u)^2 ((N-1) (s-1)-u)} 
\end{equation}
This expression provides an exact result to order $N^{-1}$. In Fig.~\ref{fig:WKB_compare} we have compared this result for different parameters to the numerical results of the full ratchet. The WKB approximation of the  mean time time to extinction in the two state model agrees  in the range $\lambda>0.5$ corresponding $\Lambda\leq1$ almost perfectly with the numerical results of the full ratchet. While the WKB-prediction is still quite good for $\Lambda=2$ it starts to deviate for
increasing values of $\Lambda$. The parameter range in which the WKB-theory works thus is slightly more restricted than in the two-state model. This can be explained by noting that for $\Lambda>1$  the two-state approximation is still valid if the ratchet  operates in the rare clicking regime, i.e.~if $NS$ is chosen to be large enough, see Fig.~\ref{fig:Moran_compare}. 
The WKB-theory on the other hand breaks down if  $x^*=e^{-\Lambda}$ is close to the absorbing state at $x=0$ independent
of $NS$.

To gain a deeper understanding of the WKB-solution one can simplify the unwieldy expression (\ref{tausolfull}) for $N\gg 1$. Expanding in $N$ and keeping only the leading order term, we obtain the approximation
\begin{equation}
	\tau = \sqrt{\frac{2\pi N}{ u}}\,\frac{s}{(s-u)^2 }\, e^{N \left(s-u
	   \left(1-\log \left(u/s\right)\right)\right)}. \label{eq:wkb_simpler}
\end{equation}
which is almost indistinguishable from the WKB-solution (\ref{tausolfull}) for $N>100$. A comparison of the exact solution 
(\ref{eq:tauMA}), the WKB-solution (\ref{tausolfull}) and the approximation (\ref{eq:wkb_simpler}) is provided in the appendix. 
It is important to note that expression (\ref{eq:wkb_simpler}) exhibits the same scaling behavior in $n^*s$ as the one found by Jain in the rare clicking regime \cite{Jain:2008uq}. To see this we rewrite this expression as 
\begin{equation}
\tau=N\,\sqrt{2\pi}\beta(\lambda) s^{-\frac{3}{2}}(n^*)^{-\frac{1}{2}} e^{n^* s \alpha(\lambda)}\, ,
\end{equation}
where $\alpha(\lambda)=(1-\lambda[1-\ln \lambda])/(1-\lambda)$ and $\beta(\lambda)=\sqrt{\lambda(1-\lambda^3)}^{-1}$.
Our solution exhibits $\lambda$-dependent functions in the exponent and the pre-factor which is in contrast to the result of Jain where these factors have to be replaced by a constant $b$. Since $\alpha(\lambda) \in [0,1]$ and $\beta(\lambda) \in [0,1]$ the values of both functions are close to the values between b=0.5 and b=0.6 which were ad hoc chosen for this constant. 
A dependence on $\Lambda$ of the factor $b$ was also observed recently by Neher and Shraiman \cite{Neher01082012}. 

\begin{figure}[htbp]
	\centering
		\includegraphics[height=3in]{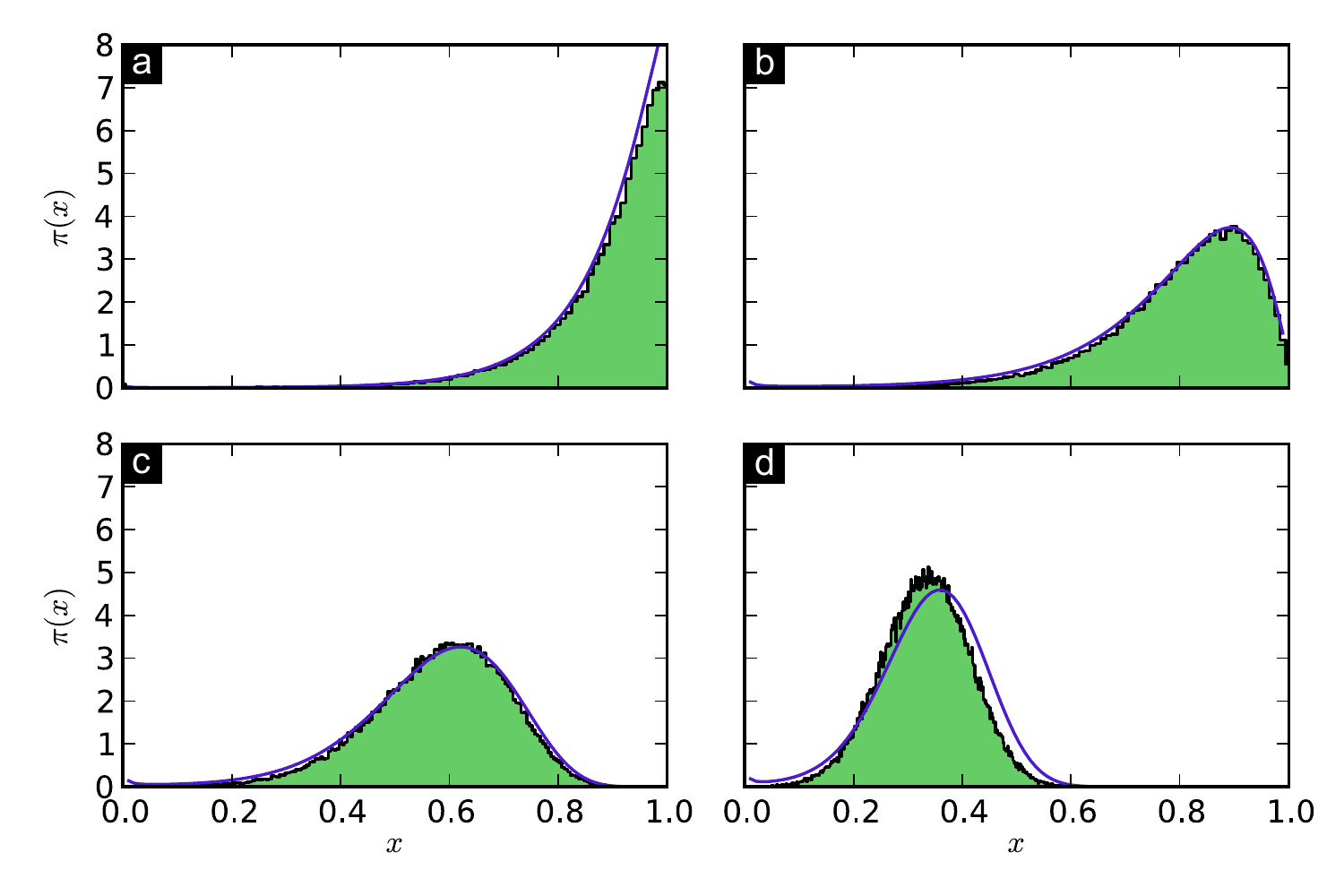}
	\caption{Comparison of the WKB-solution \eqref{QSDfullsol} for the QSD $\pi(x)$ (blue line) with the distribution of the fittest class of the full ratchet obtained by numerical simulations for $10^5$ realizations (green histogram) at $0.1\%$ of the respective click times. At this time, the distribution of the class with the lowest number of mutations has already relaxed to the quasi-stationary state, while in almost no realization a click has already occurred. The parameters used are (a) $N{=}100, \Lambda{=}0.1$, (b) $N{=}100, \Lambda{=}0.2$, (c) $N{=}200, \Lambda{=}0.5$, (d) $N{=}500, \Lambda{=}1$, and in all cases $S{=}0.1$. The analytic solution of the two-state model fits the numerical distribution obtained for the full ratchet very well. Deviations occur when the fixed point of the deterministic solution, $x^*$, begins to approach absorbing point at $x=0$, which is where the WKB approximation is expected to break down.}
	\label{fig:QSD_compare}
\end{figure}

The WKB-theory not only yields results for the ratchet rate but is also capable of describing the frequency distribution of the
fittest class in the metastable state, i.e.~the QSD, because the parameters of the two-state model were chosen such that the size of the fittest classes match in both models. One can therefore expect that also the QSD of the fittest class is approximately the same in both models. In  Fig.~\ref{fig:QSD_compare} we have compared the numerically obtained size of the fittest class in the full ratchet model 
with the WKB-solution (\ref{QSDfullsol}) and observe a striking agreement. As anticipated the WKB-theory starts to deviate if the
deterministic fixed point $x^*$ is close to the absorbing point at $x{=}0$ and if $x^*\simeq1$. 

\section{Comparison of Moran, Wright-Fisher and diffusion models}
Previous studies of Muller's ratchet have mostly considered the diffusion equation \eqref{eq:diff} to approach Muller's ratchet analytically, while numerical simulations have relied on Haigh's model with non-overlapping generations using Wright-Fisher sampling, Eq.~\eqref{Haighmodel}. Since in this work we have used a Moran formulation of the ratchet with overlapping generations, we now proceed to compare our results with the Wright-Fisher and diffusion formulations. We note that some care has to be taken to ensure that the diffusive limit of the Wright-Fisher model has the same diffusion constant as the corresponding Moran formulation, since these usually differ by a factor of two \cite{ewens2004mathematical}. Since fluctuations scale with $N^{-1/2}$, one possibility to take this into account is to consider the Wright-Fisher model with $N/2$ individuals, which is what we do in the simulations presented below.

\begin{figure}[htbp]
	\centering
		\includegraphics[height=3in]{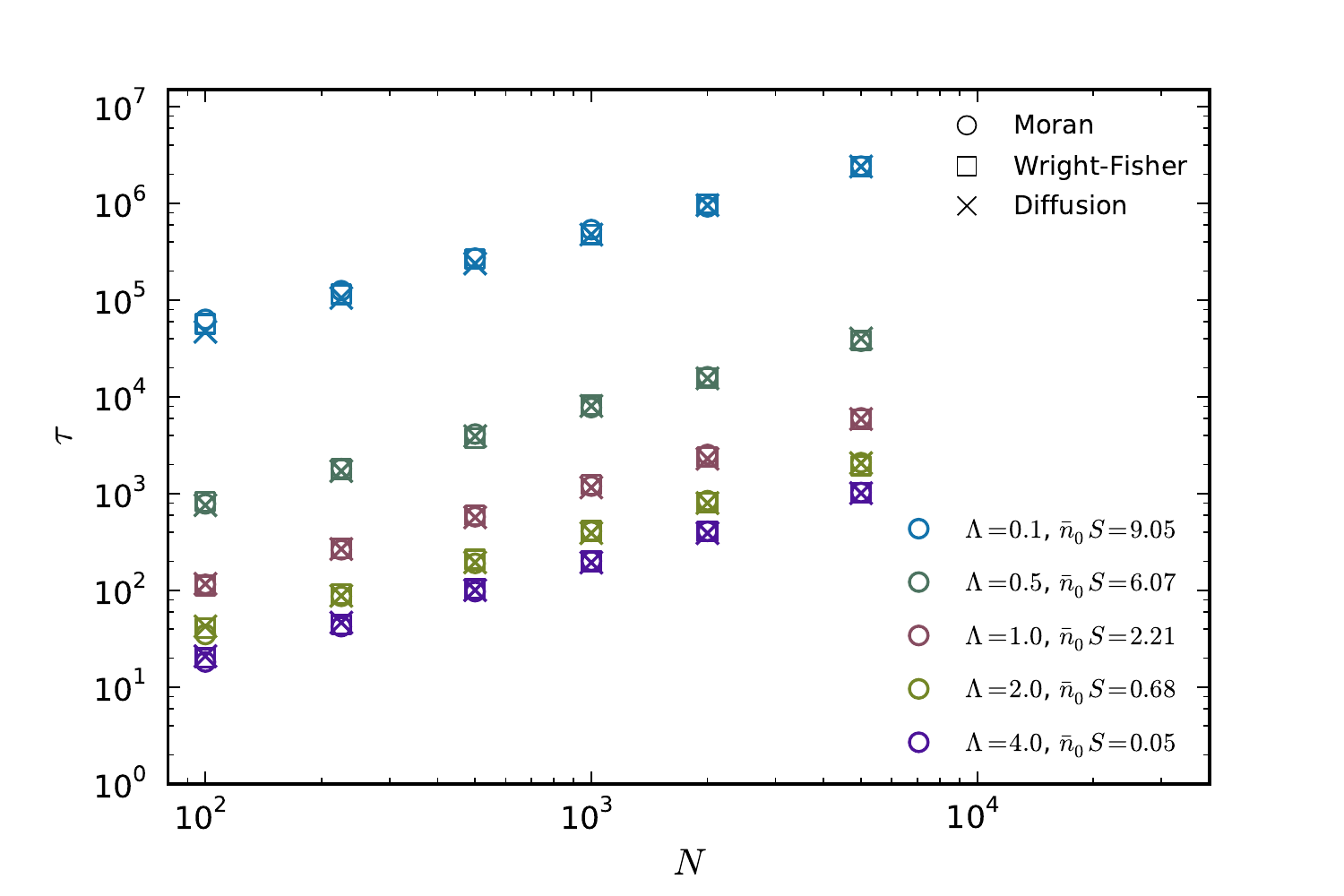}
	\caption{Comparison of the average click times $\tau$ of Muller's ratchet for the Moran model (circles), Wright-Fisher model (squares) and the diffusion limit of the two models (crosses). Same sets of parameters $U$ and $S$ are shown in the same color. We observe perfect coincidence of all three models for both slowly and fast clicking ratchets.}
	\label{fig:figure_Mor_WF}
\end{figure}

We have performed numerical simulations of the Wright-Fisher model and have numerically integrated the diffusion equation \eqref{eq:diff} using stochastic Runge-Kutta methods. To compare the three different approaches, the click times averaged over 1000 realizations for each model for different values of $\Lambda$ and $N S$ similarly to the previous sections are presented in Fig.~\ref{fig:figure_Mor_WF}.
We observe excellent agreement of the two macroscopic models and the diffusive description for slow and fast ratchets.

\section{Conclusions and Discussion}

Muller's ratchet is a model where a single rare large fluctuation influences
the fate of the whole population. Therefore a quantitative treatment of this model needs to
correctly account for such rare fluctuations which renders the analysis highly non-trivial despite
its simple formulation. In this article we have considered a Moran formulation of Muller's ratchet. 
When the ratchet clicks infrequently the population relaxes to a metastable state after each
click event. For this regime, where a click can be attributed to a rare large event, we have proposed an approximative
model which can be described  by a one-dimensional master equation. 
The exact solution of this model for the average time between successive clicks 
agrees almost perfectly with the numerical results of the full ratchet model. To gain further insight we furthermore
treated the proposed model perturbatively, employing a recently developed scheme which is particularly
well suited to the description of rare large fluctuations. We obtained a closed-form expression for the inverse ratchet rate 
which is in excellent accordance with the results obtained by numerical simulations of the full ratchet. 
It is worth pointing out that the employed WKB-theory is a non-diffusive approximation in contrast to the diffusive
approaches applied in most other works on the rate of Muller's ratchet. Additionally the WKB approach allows
for an analytical calculation of the frequency distribution of the fittest class in the quasi-stationary state. Given that
the two approximations, namely the reduction to two classes and the WKB-approach, are applied, the agreement of the
analytical result with the numerical simulation of the full ratchet model is striking. This finding, in turn, strongly supports the
validity of the reduced two-state model. To our knowledge this is the first time
that analytical expressions for the quasi-stationary distribution of the fittest class could be calculated. The expression 
(\ref{QSDfullsol}) could also be of interest for future experiments where the average time between successive ratchet
clicks is large because the ratchet relaxes very quickly to its metastable state. As we have shown this relaxation time
is only a fraction of the average inter-click-time and therefore might be experimentally more accessible in the slow ratchet
regime. Finally, we have confirmed that our results are in excellent agreement with the diffusion approximation and the corresponding Wright-Fisher formulation of the ratchet.

\section{Appendix}
\subsection{WKB-approximation and exact solution of the two-state model}
\begin{figure}[htbp]
	\centering
		\includegraphics[height=3in]{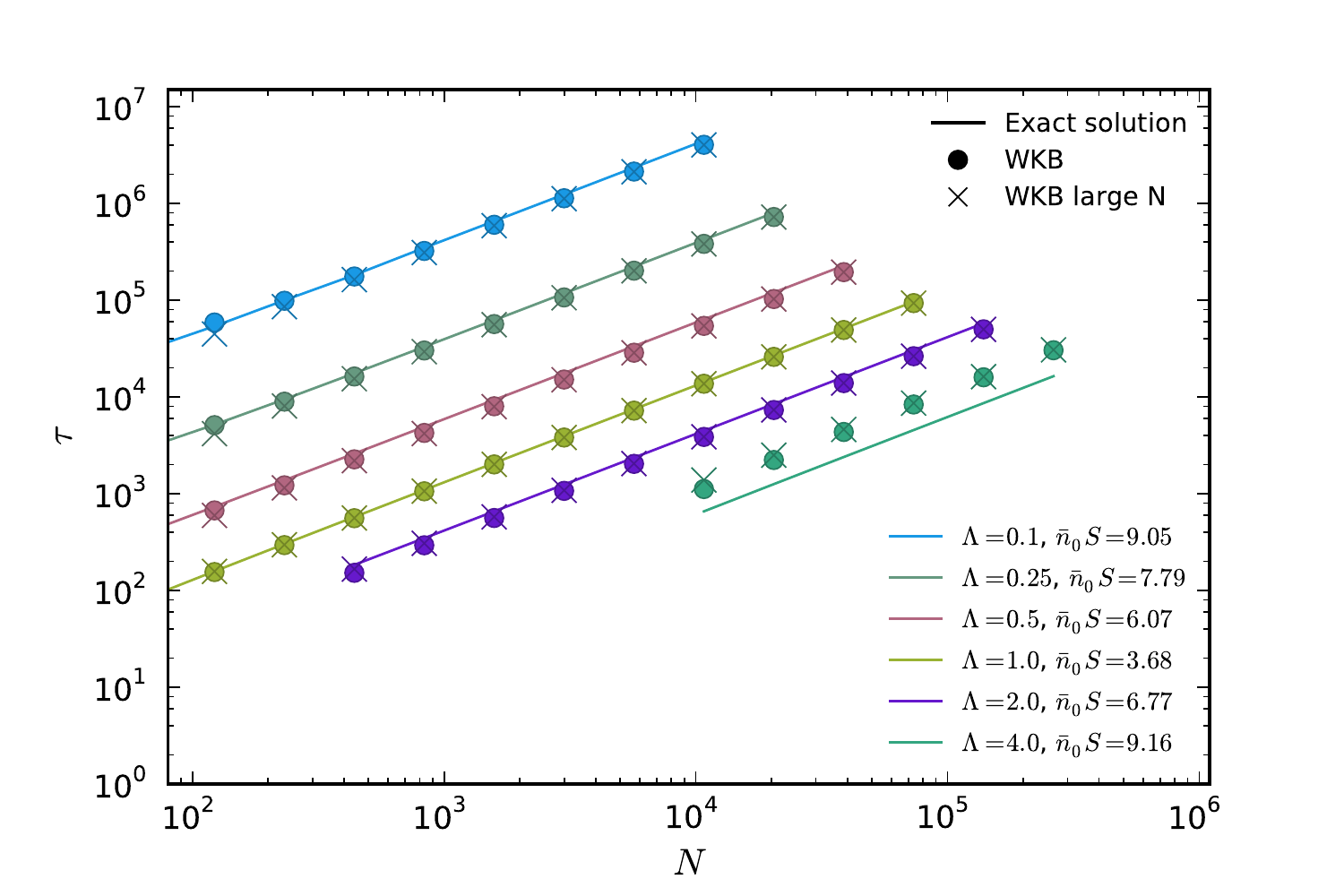}
	\caption{Comparison of the analytical click times calculated using the exact solution Eq.~\eqref{eq:tauMA}, the WKB approximation Eq.~\eqref{tausolfull}, and the simplified WKB expression for large $N$, Eq.~\eqref{eq:wkb_simpler}. }
	\label{fig:figure_App1}
\end{figure}

We compare the exact solution of the click times in the two-state model given by Eq.~\eqref{eq:tauMA} to the analytic approximation calculated from the WKB-approximation, Eq.~\eqref{tausolfull}, and to the approximation of the WKB-result for large $N$, Eq.~\eqref{eq:wkb_simpler}, in Fig.~\ref{fig:figure_App1}. The WKB results are in excellent agreement with the exact solution for the rare clicking regime when the equilibrium point of the deterministic equation, $x^*=e^{-\Lambda}$ is sufficiently far away from $0$. For large $\Lambda$, the WKB approximation begins to deviate from the exact solution. 

For the approximation of the WKB solution for large $N$, which yields the simple expression Eq.~\eqref{eq:wkb_simpler}, we observe excellent agreement with the full WKB solution when $N > 100$.


\begin{thebibliography}{41}
\expandafter\ifx\csname natexlab\endcsname\relax\def\natexlab#1{#1}\fi
\expandafter\ifx\csname bibnamefont\endcsname\relax
  \def\bibnamefont#1{#1}\fi
\expandafter\ifx\csname bibfnamefont\endcsname\relax
  \def\bibfnamefont#1{#1}\fi
\expandafter\ifx\csname citenamefont\endcsname\relax
  \def\citenamefont#1{#1}\fi
\expandafter\ifx\csname url\endcsname\relax
  \def\url#1{\texttt{#1}}\fi
\expandafter\ifx\csname urlprefix\endcsname\relax\def\urlprefix{URL }\fi
\providecommand{\bibinfo}[2]{#2}
\providecommand{\eprint}[2][]{\url{#2}}

\bibitem[{\citenamefont{Andersson and Hughes}(1996)}]{Andersson:1996uo}
\bibinfo{author}{\bibnamefont{Andersson}, \bibfnamefont{D.~I.}}, and
  \bibinfo{author}{\bibfnamefont{D.}~\bibnamefont{Hughes}},
  \bibinfo{year}{1996}, \bibinfo{journal}{Proceedings of the National Academy
  of Sciences} \textbf{\bibinfo{volume}{93}}(\bibinfo{number}{2}),
  \bibinfo{pages}{906}.

\bibitem[{\citenamefont{Assaf and Meerson}(2006)}]{Assaf:2006ii}
\bibinfo{author}{\bibnamefont{Assaf}, \bibfnamefont{M.}}, and
  \bibinfo{author}{\bibfnamefont{B.}~\bibnamefont{Meerson}},
  \bibinfo{year}{2006}, \bibinfo{journal}{Physical Review Letters}
  \textbf{\bibinfo{volume}{97}}(\bibinfo{number}{20}), \bibinfo{pages}{200602}.

\bibitem[{\citenamefont{Assaf and Meerson}(2010)}]{Assaf:2010de}
\bibinfo{author}{\bibnamefont{Assaf}, \bibfnamefont{M.}}, and
  \bibinfo{author}{\bibfnamefont{B.}~\bibnamefont{Meerson}},
  \bibinfo{year}{2010}, \bibinfo{journal}{Physical Review E}
  \textbf{\bibinfo{volume}{81}}(\bibinfo{number}{2}), \bibinfo{pages}{021116}.

\bibitem[{\citenamefont{Assaf and Mobilia}(2011)}]{Assaf:2011hs}
\bibinfo{author}{\bibnamefont{Assaf}, \bibfnamefont{M.}}, and
  \bibinfo{author}{\bibfnamefont{M.}~\bibnamefont{Mobilia}},
  \bibinfo{year}{2011}, \bibinfo{journal}{Journal of Theoretical Biology}
  \textbf{\bibinfo{volume}{275}}(\bibinfo{number}{1}), \bibinfo{pages}{93}.

\bibitem[{\citenamefont{Assaf} \emph{et~al.}(2011)\citenamefont{Assaf, Roberts,
  and Luthey-Schulten}}]{Assaf:2011iw}
\bibinfo{author}{\bibnamefont{Assaf}, \bibfnamefont{M.}},
  \bibinfo{author}{\bibfnamefont{E.}~\bibnamefont{Roberts}}, and
  \bibinfo{author}{\bibfnamefont{Z.}~\bibnamefont{Luthey-Schulten}},
  \bibinfo{year}{2011}, \bibinfo{journal}{Physical Review Letters}
  \textbf{\bibinfo{volume}{106}}(\bibinfo{number}{24}),
  \bibinfo{pages}{248102}.

\bibitem[{\citenamefont{Barton and Charlesworth}(1998)}]{Barton:1998uq}
\bibinfo{author}{\bibnamefont{Barton}, \bibfnamefont{N.~H.}}, and
  \bibinfo{author}{\bibfnamefont{B.}~\bibnamefont{Charlesworth}},
  \bibinfo{year}{1998}, \bibinfo{journal}{Science}
  \textbf{\bibinfo{volume}{281}}(\bibinfo{number}{5385}),
  \bibinfo{pages}{1986}.

\bibitem[{\citenamefont{Black and McKane}(2011)}]{Black:2011ep}
\bibinfo{author}{\bibnamefont{Black}, \bibfnamefont{A.~J.}}, and
  \bibinfo{author}{\bibfnamefont{A.~J.} \bibnamefont{McKane}},
  \bibinfo{year}{2011}, \bibinfo{journal}{Journal of Statistical Mechanics:
  Theory and Experiment} \textbf{\bibinfo{volume}{2011}}(\bibinfo{number}{12}),
  \bibinfo{pages}{P12006}.

\bibitem[{\citenamefont{Black} \emph{et~al.}(2012)\citenamefont{Black,
  Traulsen, and Galla}}]{Black:2012dz}
\bibinfo{author}{\bibnamefont{Black}, \bibfnamefont{A.~J.}},
  \bibinfo{author}{\bibfnamefont{A.}~\bibnamefont{Traulsen}}, and
  \bibinfo{author}{\bibfnamefont{T.}~\bibnamefont{Galla}},
  \bibinfo{year}{2012}, \bibinfo{journal}{Physical Review Letters}
  \textbf{\bibinfo{volume}{109}}(\bibinfo{number}{2}), \bibinfo{pages}{028101}.

\bibitem[{\citenamefont{Blythe and McKane}(2007)}]{blythe_stochastic_2007}
\bibinfo{author}{\bibnamefont{Blythe}, \bibfnamefont{R.~A.}}, and
  \bibinfo{author}{\bibfnamefont{A.~J.} \bibnamefont{McKane}},
  \bibinfo{year}{2007}, \bibinfo{journal}{Journal of Statistical Mechanics:
  Theory and Experiment} \textbf{\bibinfo{volume}{2007}}(\bibinfo{number}{07}),
  \bibinfo{pages}{P07018}.

\bibitem[{\citenamefont{Chao}(1990)}]{Chao:1990bx}
\bibinfo{author}{\bibnamefont{Chao}, \bibfnamefont{L.}}, \bibinfo{year}{1990},
  \bibinfo{journal}{Nature}
  \textbf{\bibinfo{volume}{348}}(\bibinfo{number}{6300}), \bibinfo{pages}{454}.

\bibitem[{\citenamefont{De~Visser and Elena}(2007)}]{DeVisser:2007gz}
\bibinfo{author}{\bibnamefont{De~Visser}, \bibfnamefont{J.~A. G.~M.}}, and
  \bibinfo{author}{\bibfnamefont{S.~F.} \bibnamefont{Elena}},
  \bibinfo{year}{2007}, \bibinfo{journal}{Nature Reviews Genetics}
  \textbf{\bibinfo{volume}{8}}(\bibinfo{number}{2}), \bibinfo{pages}{139}.

\bibitem[{\citenamefont{Doering} \emph{et~al.}(2006)\citenamefont{Doering,
  Sargsyan, and Sander}}]{Doering:2013bd}
\bibinfo{author}{\bibnamefont{Doering}, \bibfnamefont{C.}},
  \bibinfo{author}{\bibfnamefont{K.}~\bibnamefont{Sargsyan}}, and
  \bibinfo{author}{\bibfnamefont{L.}~\bibnamefont{Sander}},
  \bibinfo{year}{2006}, \bibinfo{journal}{Multiscale Modeling {\&} Simulation}
  \textbf{\bibinfo{volume}{3}}(\bibinfo{number}{2}), \bibinfo{pages}{283}.

\bibitem[{\citenamefont{Duarte} \emph{et~al.}(1992)\citenamefont{Duarte,
  Clarke, Moya, Domingo, and Holland}}]{Duarte:1992vj}
\bibinfo{author}{\bibnamefont{Duarte}, \bibfnamefont{E.}},
  \bibinfo{author}{\bibfnamefont{D.}~\bibnamefont{Clarke}},
  \bibinfo{author}{\bibfnamefont{A.}~\bibnamefont{Moya}},
  \bibinfo{author}{\bibfnamefont{E.}~\bibnamefont{Domingo}}, and
  \bibinfo{author}{\bibfnamefont{J.}~\bibnamefont{Holland}},
  \bibinfo{year}{1992}, \bibinfo{journal}{Proceedings of the National Academy
  of Sciences} \textbf{\bibinfo{volume}{89}}(\bibinfo{number}{13}),
  \bibinfo{pages}{6015}.

\bibitem[{\citenamefont{Dykman} \emph{et~al.}(2008)\citenamefont{Dykman,
  Schwartz, and Landsman}}]{Dykman:2008ig}
\bibinfo{author}{\bibnamefont{Dykman}, \bibfnamefont{M.}},
  \bibinfo{author}{\bibfnamefont{I.}~\bibnamefont{Schwartz}}, and
  \bibinfo{author}{\bibfnamefont{A.}~\bibnamefont{Landsman}},
  \bibinfo{year}{2008}, \bibinfo{journal}{Physical Review Letters}
  \textbf{\bibinfo{volume}{101}}(\bibinfo{number}{7}), \bibinfo{pages}{078101}.

\bibitem[{\citenamefont{Dykman} \emph{et~al.}(1994)\citenamefont{Dykman, Mori,
  Ross, and Hunt}}]{Dykman:1994hw}
\bibinfo{author}{\bibnamefont{Dykman}, \bibfnamefont{M.~I.}},
  \bibinfo{author}{\bibfnamefont{E.}~\bibnamefont{Mori}},
  \bibinfo{author}{\bibfnamefont{J.}~\bibnamefont{Ross}}, and
  \bibinfo{author}{\bibfnamefont{P.~M.} \bibnamefont{Hunt}},
  \bibinfo{year}{1994}, \bibinfo{journal}{The Journal of Chemical Physics}
  \textbf{\bibinfo{volume}{100}}(\bibinfo{number}{8}), \bibinfo{pages}{5735}.

\bibitem[{\citenamefont{Etheridge} \emph{et~al.}(2009)\citenamefont{Etheridge,
  Pfaffelhuber, and Wakolbinger}}]{Etheridge:2007tv}
\bibinfo{author}{\bibnamefont{Etheridge}, \bibfnamefont{A.}},
  \bibinfo{author}{\bibfnamefont{P.}~\bibnamefont{Pfaffelhuber}}, and
  \bibinfo{author}{\bibfnamefont{A.}~\bibnamefont{Wakolbinger}},
  \bibinfo{year}{2009}, in \emph{\bibinfo{booktitle}{Trends in Stochastic
  Analysis}}, edited by
  \bibinfo{editor}{\bibfnamefont{J.}~\bibnamefont{Blath}},
  \bibinfo{editor}{\bibfnamefont{P.}~\bibnamefont{M{\"o}rters}}, and
  \bibinfo{editor}{\bibfnamefont{M.}~\bibnamefont{Scheutzow}}
  (\bibinfo{publisher}{Cambridge University Press},
  \bibinfo{address}{Cambridge}), pp. \bibinfo{pages}{365--390}.

\bibitem[{\citenamefont{Ewens}(2004)}]{ewens2004mathematical}
\bibinfo{author}{\bibnamefont{Ewens}, \bibfnamefont{W.~J.}},
  \bibinfo{year}{2004}, \emph{\bibinfo{title}{{Mathematical population
  genetics: I. Theoretical introduction}}}, volume~\bibinfo{volume}{27}
  (\bibinfo{publisher}{Springer}).

\bibitem[{\citenamefont{Felsenstein}(1974)}]{Felsenstein:1974wq}
\bibinfo{author}{\bibnamefont{Felsenstein}, \bibfnamefont{J.}},
  \bibinfo{year}{1974}, \bibinfo{journal}{Genetics}
  \textbf{\bibinfo{volume}{78}}(\bibinfo{number}{2}), \bibinfo{pages}{737}.

\bibitem[{\citenamefont{Gardiner}(2009)}]{gardiner2009stochastic}
\bibinfo{author}{\bibnamefont{Gardiner}, \bibfnamefont{C.}},
  \bibinfo{year}{2009}, \emph{\bibinfo{title}{{Stochastic Methods: A Handbook
  for the Natural and Social Sciences}}}, Springer Series in Synergetics
  (\bibinfo{publisher}{Springer}).

\bibitem[{\citenamefont{Gessler}(1995)}]{Gessler:1995jh}
\bibinfo{author}{\bibnamefont{Gessler}, \bibfnamefont{D.~D.~G.}},
  \bibinfo{year}{1995}, \bibinfo{journal}{Genetical Research}
  \textbf{\bibinfo{volume}{66}}(\bibinfo{number}{03}), \bibinfo{pages}{241}.

\bibitem[{\citenamefont{Gordo and Charlesworth}(2000)}]{Gordo01032000}
\bibinfo{author}{\bibnamefont{Gordo}, \bibfnamefont{I.}}, and
  \bibinfo{author}{\bibfnamefont{B.}~\bibnamefont{Charlesworth}},
  \bibinfo{year}{2000}, \bibinfo{journal}{Genetics}
  \textbf{\bibinfo{volume}{154}}(\bibinfo{number}{3}), \bibinfo{pages}{1379}.

\bibitem[{\citenamefont{Haigh}(1978)}]{Haigh1978251}
\bibinfo{author}{\bibnamefont{Haigh}, \bibfnamefont{J.}}, \bibinfo{year}{1978},
  \bibinfo{journal}{Theoretical Population Biology}
  \textbf{\bibinfo{volume}{14}}(\bibinfo{number}{2}), \bibinfo{pages}{251}.

\bibitem[{\citenamefont{Hanggi} \emph{et~al.}(1984)\citenamefont{Hanggi,
  Grabert, Talkner, and Thomas}}]{Hanggi:1984bo}
\bibinfo{author}{\bibnamefont{Hanggi}, \bibfnamefont{P.}},
  \bibinfo{author}{\bibfnamefont{H.}~\bibnamefont{Grabert}},
  \bibinfo{author}{\bibfnamefont{P.}~\bibnamefont{Talkner}}, and
  \bibinfo{author}{\bibfnamefont{H.}~\bibnamefont{Thomas}},
  \bibinfo{year}{1984}, \bibinfo{journal}{Physical Review A}
  \textbf{\bibinfo{volume}{29}}(\bibinfo{number}{1}), \bibinfo{pages}{371}.

\bibitem[{\citenamefont{Higgs and Woodcock}(1995)}]{Higgs:1995dt}
\bibinfo{author}{\bibnamefont{Higgs}, \bibfnamefont{P.}}, and
  \bibinfo{author}{\bibfnamefont{G.}~\bibnamefont{Woodcock}},
  \bibinfo{year}{1995}, \bibinfo{journal}{Journal of mathematical biology}
  \textbf{\bibinfo{volume}{33}}(\bibinfo{number}{7}), \bibinfo{pages}{677}.

\bibitem[{\citenamefont{Howe and Denver}(2008)}]{Howe:2008fx}
\bibinfo{author}{\bibnamefont{Howe}, \bibfnamefont{D.~K.}}, and
  \bibinfo{author}{\bibfnamefont{D.~R.} \bibnamefont{Denver}},
  \bibinfo{year}{2008}, \bibinfo{journal}{BMC Evolutionary Biology}
  \textbf{\bibinfo{volume}{8}}(\bibinfo{number}{1}), \bibinfo{pages}{62}.

\bibitem[{\citenamefont{Jain}(2008)}]{Jain:2008uq}
\bibinfo{author}{\bibnamefont{Jain}, \bibfnamefont{K.}}, \bibinfo{year}{2008},
  \bibinfo{journal}{Genetics}
  \textbf{\bibinfo{volume}{179}}(\bibinfo{number}{4}), \bibinfo{pages}{2125}.

\bibitem[{\citenamefont{Lynch}(1996)}]{Lynch:1996uf}
\bibinfo{author}{\bibnamefont{Lynch}, \bibfnamefont{M.}}, \bibinfo{year}{1996},
  \bibinfo{journal}{Molecular biology and evolution}
  \textbf{\bibinfo{volume}{13}}(\bibinfo{number}{1}), \bibinfo{pages}{209}.

\bibitem[{\citenamefont{Lynch} \emph{et~al.}(1993)\citenamefont{Lynch,
  B{\"u}rger, Butcher, and Gabriel}}]{Lynch:1993tl}
\bibinfo{author}{\bibnamefont{Lynch}, \bibfnamefont{M.}},
  \bibinfo{author}{\bibfnamefont{R.}~\bibnamefont{B{\"u}rger}},
  \bibinfo{author}{\bibfnamefont{D.}~\bibnamefont{Butcher}}, and
  \bibinfo{author}{\bibfnamefont{W.}~\bibnamefont{Gabriel}},
  \bibinfo{year}{1993}, \bibinfo{journal}{Journal of Heredity}
  \textbf{\bibinfo{volume}{84}}(\bibinfo{number}{5}), \bibinfo{pages}{339}.

\bibitem[{\citenamefont{Muller}(1964)}]{Muller19642}
\bibinfo{author}{\bibnamefont{Muller}, \bibfnamefont{H.~J.}},
  \bibinfo{year}{1964}, \bibinfo{journal}{Mutation Research/Fundamental and
  Molecular Mechanisms of Mutagenesis}
  \textbf{\bibinfo{volume}{1}}(\bibinfo{number}{1}), \bibinfo{pages}{2}.

\bibitem[{\citenamefont{Neher and Shraiman}(2012)}]{Neher01082012}
\bibinfo{author}{\bibnamefont{Neher}, \bibfnamefont{R.~A.}}, and
  \bibinfo{author}{\bibfnamefont{B.~I.} \bibnamefont{Shraiman}},
  \bibinfo{year}{2012}, \bibinfo{journal}{Genetics}
  \textbf{\bibinfo{volume}{191}}(\bibinfo{number}{4}), \bibinfo{pages}{1283}.

\bibitem[{\citenamefont{Ovaskainen and Meerson}(2010)}]{Ovaskainen:2010cl}
\bibinfo{author}{\bibnamefont{Ovaskainen}, \bibfnamefont{O.}}, and
  \bibinfo{author}{\bibfnamefont{B.}~\bibnamefont{Meerson}},
  \bibinfo{year}{2010}, \bibinfo{journal}{Trends in Ecology {\&} Evolution}
  \textbf{\bibinfo{volume}{25}}(\bibinfo{number}{11}), \bibinfo{pages}{643}.

\bibitem[{\citenamefont{Park} \emph{et~al.}(2010)\citenamefont{Park, Simon, and
  Krug}}]{park_speed_2010}
\bibinfo{author}{\bibnamefont{Park}, \bibfnamefont{S.-C.}},
  \bibinfo{author}{\bibfnamefont{D.}~\bibnamefont{Simon}}, and
  \bibinfo{author}{\bibfnamefont{J.}~\bibnamefont{Krug}}, \bibinfo{year}{2010},
  \bibinfo{journal}{Journal of Statistical Physics}
  \textbf{\bibinfo{volume}{138}}(\bibinfo{number}{1-3}), \bibinfo{pages}{381}.

\bibitem[{\citenamefont{Rice}(1987)}]{Rice:1987wo}
\bibinfo{author}{\bibnamefont{Rice}, \bibfnamefont{W.~R.}},
  \bibinfo{year}{1987}, \bibinfo{journal}{Genetics}
  \textbf{\bibinfo{volume}{116}}(\bibinfo{number}{1}), \bibinfo{pages}{161}.

\bibitem[{\citenamefont{Rice}(1994)}]{Rice:1994wt}
\bibinfo{author}{\bibnamefont{Rice}, \bibfnamefont{W.~R.}},
  \bibinfo{year}{1994}, \bibinfo{journal}{Science}
  \textbf{\bibinfo{volume}{263}}, \bibinfo{pages}{230}.

\bibitem[{\citenamefont{R{\"o}{\ss}ler}(2010)}]{Roessler:2010:RMS:1958393.1958409}
\bibinfo{author}{\bibnamefont{R{\"o}{\ss}ler}, \bibfnamefont{A.}},
  \bibinfo{year}{2010}, \bibinfo{journal}{SIAM J. Numer. Anal.}
  \textbf{\bibinfo{volume}{48}}(\bibinfo{number}{3}), \bibinfo{pages}{922}.

\bibitem[{\citenamefont{Rouzine} \emph{et~al.}(2008)\citenamefont{Rouzine,
  Brunet, and Wilke}}]{Rouzine200824}
\bibinfo{author}{\bibnamefont{Rouzine}, \bibfnamefont{I.~M.}},
  \bibinfo{author}{\bibfnamefont{{\'E}.}~\bibnamefont{Brunet}}, and
  \bibinfo{author}{\bibfnamefont{C.~O.} \bibnamefont{Wilke}},
  \bibinfo{year}{2008}, \bibinfo{journal}{Theoretical Population Biology}
  \textbf{\bibinfo{volume}{73}}(\bibinfo{number}{1}), \bibinfo{pages}{24}.

\bibitem[{\citenamefont{Schwartz} \emph{et~al.}(2011)\citenamefont{Schwartz,
  Forgoston, Bianco, and Shaw}}]{Schwartz:2011uy}
\bibinfo{author}{\bibnamefont{Schwartz}, \bibfnamefont{I.~B.}},
  \bibinfo{author}{\bibfnamefont{E.}~\bibnamefont{Forgoston}},
  \bibinfo{author}{\bibfnamefont{S.}~\bibnamefont{Bianco}}, and
  \bibinfo{author}{\bibfnamefont{L.~B.} \bibnamefont{Shaw}},
  \bibinfo{year}{2011}, \bibinfo{journal}{Journal of The Royal Society
  Interface} \textbf{\bibinfo{volume}{8}}(\bibinfo{number}{65}),
  \bibinfo{pages}{1699}.

\bibitem[{\citenamefont{Stephan} \emph{et~al.}(1993)\citenamefont{Stephan,
  Chao, and Smale}}]{Stephan:1993do}
\bibinfo{author}{\bibnamefont{Stephan}, \bibfnamefont{W.}},
  \bibinfo{author}{\bibfnamefont{L.}~\bibnamefont{Chao}}, and
  \bibinfo{author}{\bibfnamefont{J.~G.} \bibnamefont{Smale}},
  \bibinfo{year}{1993}, \bibinfo{journal}{Genetical Research}
  \textbf{\bibinfo{volume}{61}}(\bibinfo{number}{03}), \bibinfo{pages}{225}.

\bibitem[{\citenamefont{Stephan and Kim}(2002)}]{StephanKim2002}
\bibinfo{author}{\bibnamefont{Stephan}, \bibfnamefont{W.}}, and
  \bibinfo{author}{\bibfnamefont{Y.}~\bibnamefont{Kim}}, \bibinfo{year}{2002},
  \emph{\bibinfo{title}{{Recent Applications of Diffusion Theory to Population
  Genetics}}}, Modern Developments in Theoretical Population Genetics
  (\bibinfo{publisher}{Oxford University Press, Oxford, UK}).

\bibitem[{\citenamefont{Waxman and Loewe}(2010)}]{Waxman:2010fb}
\bibinfo{author}{\bibnamefont{Waxman}, \bibfnamefont{D.}}, and
  \bibinfo{author}{\bibfnamefont{L.}~\bibnamefont{Loewe}},
  \bibinfo{year}{2010}, \bibinfo{journal}{Journal of Theoretical Biology}
  \textbf{\bibinfo{volume}{264}}(\bibinfo{number}{4}), \bibinfo{pages}{1120}.

\bibitem[{\citenamefont{Zeyl} \emph{et~al.}(2001)\citenamefont{Zeyl, Mizesko,
  and De~Visser}}]{Zeyl:2001if}
\bibinfo{author}{\bibnamefont{Zeyl}, \bibfnamefont{C.}},
  \bibinfo{author}{\bibfnamefont{M.}~\bibnamefont{Mizesko}}, and
  \bibinfo{author}{\bibfnamefont{J.~A. G.~M.} \bibnamefont{De~Visser}},
  \bibinfo{year}{2001}, \bibinfo{journal}{Evolution}
  \textbf{\bibinfo{volume}{55}}(\bibinfo{number}{5}), \bibinfo{pages}{909}.

\end{thebibliography}
\end{document}